\documentclass[PRD,preprintnumbers,floatfix,aps,notitlepage,nofootinbib,twocolumn,showpacs,amssymb]{revtex4-1}
\usepackage{amsmath,amsfonts,bm}
\usepackage{graphicx}
\usepackage{cancel}
\usepackage[colorlinks, linkcolor={red},citecolor={blue}]{hyperref}
\usepackage{xcolor}

\newcommand{\rh}{r_{\rm h}}

\begin{document}
\title{Regular hairy black holes through Minkowski deformation}
%
\author{Jorge Ovalle}
\email{jorge.ovalle@physics.slu.cz}
\affiliation{Research Centre for Theoretical Physics and Astrophysics,
Institute of Physics, Silesian University in Opava, CZ-746 01 Opava, Czech Republic.
\\
Universidad Central de Chile, Vicerrector\'ia Acad\'emica, Toesca 1783, Santiago, Chile}
\author{Roberto Casadio}
\email{casadio@bo.infn.it}
\affiliation{Dipartimento di Fisica e Astronomia ``A.~Righi'',
	Universit\`a di Bologna,
	40126 Bologna, Italy
	\\
	Istituto Nazionale di Fisica Nucleare, I.S.~FLAG,
	Sezione di Bologna, 40127 Bologna, Italy}
\author{Andrea~Giusti}
\email{agiusti@phys.ethz.ch}
\affiliation{Institute for Theoretical Physics, ETH Zurich, Wolfgang-Pauli-Strasse 27, 8093 Zurich, Switzerland}
\begin{abstract}
Static and stationary regular black holes are examined under a minimal set of requirements consisting of
(i) the existence of a well defined event horizon and
(ii) the weak energy condition for matter sourcing the geometry.
We perform our analysis by means of the gravitational decoupling approach and find hairy solutions free
of curvature singularities.
We identify the matter source producing a deformation of the Minkowski vacuum such that
the maximum deformation is the Schwarzschild solution for the static case, and the Kerr metric
for the stationary case.
\end{abstract} 
\maketitle
%
\section{Introduction}
Possible conditions for circumventing the {\it no-hair\/} conjecture have been investigated
for a long time and in different
scenarios~\cite{Martinez:2004nb,Sotiriou:2011dz,Babichev:2013cya,Sotiriou:2013qea,Antoniou:2017acq,Antoniou:2017hxj}.
An option is to fill the would-be static vacuum of General Relativity (GR) with a source, 
possibly of fundamental origin, which is often described using a scalar field~\cite{Herdeiro:2015waa}.
One of the main reasons to do so is to eliminate the singularities which should form at the end of the
gravitational collapse according to GR.
Even though the Cosmic Censorship Conjecture (CCC) states that these singularities are always
hidden inside an event horizon~\cite{Penrose:1969pc,Hawking:1973uf},
their very prediction should be taken as a clear signal about the limitations of the theory.
\par
Regarding cancellation of singularities by hairy solutions, a plethora of new regular
black holes (BHs) have been proposed in recent years.
There is a relatively simple way to interpret the matter source used to evade singularities
in terms of nonlinear electrodynamics~\cite{Salazar:1987ap,Ayon-Beato:1998hmi}
(see also Refs.~\cite{Bronnikov:2000vy,Dymnikova:2004zc,Balart:2014cga,Toshmatov:2014nya,Fan:2016hvf}).
Unfortunately, classically regular solutions usually contain a Cauchy horizon, a null hypersurface
beyond which predictability breaks down~\cite{Poisson:1989zz,Poisson:1990eh},
and which turn out to be quite problematic~\cite{Bonanno:2020fgp,
Carballo-Rubio:2022kad,Franzin:2022wai,Bonanno:2022jjp}.
A first step in the direction of avoiding such issues could be to describe matter in the most
general way possible, ensuring a flexible enough scenario with the least number of restrictions.
This is precisely the scheme followed in this work, where the Schwarzschild vacuum
is filled with a generic static and spherically symmetric source $\theta_{\mu\nu}$ which we call
a ``tensor-vacuum''. 
This scheme is a direct consequence of the gravitational decoupling (GD)
method~\cite{Ovalle:2017fgl,Ovalle:2019qyi}, and has resulted particularly useful
to generate hairy BHs in both the spherically symmetric~\cite{Ovalle:2018umz,Ovalle:2020kpd}
and axially symmetric case~\cite{Contreras:2021yxe}
(see also Ref.~\cite{Islam:2021dyk,daRocha:2020gee,Ovalle:2021jzf,Afrin:2021imp,Ramos:2021jta,
Meert:2021khi,Mahapatra:2022xea,Cavalcanti:2022cga,Omwoyo:2021uah,Mahapatra:2022xea,Avalos:2023jeh,Avalos:2023ywb,Wu:2023wld}).
One of the most attractive features of this scheme, is that it allows to introduce
a minimal set of requirements, without jeopardizing the static or stationary vacuum
far from the source.
In this respect, the goal of this work is to find regular BHs for the static and rotational cases,
which are asymptotically flat and satisfy some of the energy conditions.
\par
The paper is organised as follows:
in Section~\ref{sec2}, we briefly review the GD scheme, showing the decoupling
of two gravitational sources for the spherically symmetric case;
in Section~\ref{sec3}, we implement the GD to produce regular hairy black holes
satisfying the weak energy condition;
in Section~\ref{sec4} we generate the axially symmetric version of the regular hairy BH;
finally, we summarize our conclusions in Section~\ref{con}.
\section{Gravitational Decoupling}
\label{sec2}
In order to be as self-contained as possible, in this Section we briefly review the GD for spherically symmetric gravitational systems
described in detail in Ref.~\cite{Ovalle:2019qyi}.
For the axially symmetric case, see Ref.~\cite{Contreras:2021yxe}.
\par
We start from the Einstein-Hilbert action
\begin{equation}
	\label{action}
	S=\int\left[\frac{R}{2\,\kappa}+{\cal L}_{\rm M}+{\cal L}_{\Theta}\right]\sqrt{-g}\,d^4x
	\ ,
\end{equation}
where $R$ is the Ricci scalar, ${\cal L}_{\rm M}$ contains standard matter fields and
${\cal L}_{\Theta}$ is a second Lagrangian density which may describe matter or be
related with a new gravitational sector beyond general relativity.
For both sources, the energy-momentum tensors are defined as usual by
\begin{eqnarray}
	&&
	T_{\mu\nu}
	=
	-\frac{2}{\sqrt{-g}}\frac{\delta\,(\sqrt{-g}\,{\cal L}_{\rm M})}{\delta\,g^{\mu\nu}}
	=
	-2\,\frac{\delta\,{\cal L}_{\rm M}}{\delta\,g^{\mu\nu}}
	+g_{\mu\nu}{\cal L}_{\rm M}
	\ ,
	\\
	&&
	\theta_{\mu\nu}
	=
	-\frac{2}{\sqrt{-g}}\frac{\delta\,(\sqrt{-g}\,{\cal L}_{\Theta})}{\delta\,g^{\mu\nu}}
	=
	-2\,\frac{\delta\,{\cal L}_{\Theta}}{\delta\,g^{\mu\nu}}
	+g_{\mu\nu}{\cal L}_{\Theta}
	\ ,
\end{eqnarray}
so that the action in Eq.~\eqref{action} yields the Einstein field equations~\footnote{We use units with $c=1$
and $\kappa=8\,\pi\,G_{\rm N}$, where $G_{\rm N}$ is Newton's constant, and signature $(-,+,+,+)$.}
\begin{equation}
	\label{corr2}
	G_{\mu\nu}
	\equiv
	R_{\mu\nu}-\frac{1}{2}\,R\, g_{\mu\nu}
	=
	\kappa\,\tilde{T}_{\mu\nu}
	\ ,
\end{equation}
with a total energy-momentum tensor given by
\begin{equation}
	\label{emt}
	\tilde{T}_{\mu\nu}
	=
	T^{\rm}_{\mu\nu}
	+
	\theta_{\mu\nu}
	\ ,
\end{equation}
which must be covariantly conserved,
\begin{equation}
	\nabla_\mu\,\tilde{T}^{\mu\nu}=0
	\ ,
	\label{dT0}
\end{equation} 
as a consequence of the Bianchi identity.
\par
For spherically symmetric and static systems, we can write the metric $g_{\mu\nu}$ as
\begin{equation}
	ds^{2}
	=
	-e^{\nu (r)}\,dt^{2}+e^{\lambda (r)}\,dr^{2}
	+r^{2}\,d\Omega^2
	\ ,
	\label{metric}
\end{equation}
where $\nu =\nu (r)$ and $\lambda =\lambda (r)$ are functions of the areal
radius $r$ only and $d\Omega^2=d\theta^{2}+\sin^{2}\theta \,d\phi ^{2}$.
The Einstein equations~(\ref{corr2}) then read
\begin{eqnarray}
	\label{ec1}
	\kappa\,
	\tilde T_0^{\ 0}
	&=&
	-\frac 1{r^2}
	+
	e^{-\lambda }\left( \frac1{r^2}-\frac{\lambda'}r\right)
	\\
	\label{ec2}
	\kappa\,
	\tilde T_1^{\ 1}
	&=&
	-\frac 1{r^2}
	+
	e^{-\lambda }\left( \frac 1{r^2}+\frac{\nu'}r\right)
	\\
	\label{ec3}
	\kappa\,
	\tilde T_2^{\ 2}
	&=&
	\frac {e^{-\lambda }}{4}
	\left(2\nu''+\nu'^2-\lambda'\nu'
	+2\,\frac{\nu'-\lambda'}r\right)
	\ ,
\end{eqnarray}
where $f'\equiv \partial_r f$ and $\tilde{T}_3^{{\ 3}}=\tilde{T}_2^{\ 2}$ due to the spherical symmetry.
By simple inspection, we can identify in Eqs.~\eqref{ec1}-\eqref{ec3} an effective energy density  
\begin{equation}
	\tilde{\epsilon}
	=-
	T_0^{\ 0}
	-
	\theta_0^{\ 0}=\epsilon+{\cal E}
	\ ,
	\label{efecden}
\end{equation}
an effective radial pressure
\begin{equation}
	\tilde{p}_{r}
	=
	T_1^{\ 1}
	+\theta_1^{\ 1}=p_r+{\cal P}_r
	\ ,
	\label{efecprera}
\end{equation}
and an effective tangential pressure
\begin{equation}
	\tilde{p}_{t}
	=
	T_2^{\ 2}
	+\theta_2^{\ 2}=p_\theta+{\cal P}_\theta
	\ ,
	\label{efecpretan}
\end{equation} 
where we clearly have 
\begin{eqnarray}
	&&
	T_\mu^{\,\,\nu}={\rm diag}\left[-\epsilon,\,p_r,\,p_\theta,\,p_\theta\right]
	\ ,
	\\
	&&
	\theta_\mu^{\,\,\nu}={\rm diag}\left[-{\cal E},\,{\cal P}_r,\,{\cal P}_\theta,\,{\cal P}_\theta\right]
	\ .
\end{eqnarray}
In general, $\Pi\equiv\tilde{p}_{\theta}-\tilde{p}_{r}$ does not vanish and the system
of Eqs.~\eqref{ec1}-\eqref{ec3} describes an anisotropic fluid.
\par
We next consider a solution to the Eqs.~\eqref{corr2} for the seed source $T_{\mu\nu}$
alone, 
which we write as
\begin{equation}
	ds^{2}
	=
	-e^{\xi (r)}\,dt^{2}
	+e^{\mu (r)}\,dr^{2}
	+
	r^{2}\,d\Omega^2
	\ ,
	\label{pfmetric}
\end{equation}
where 
\begin{equation}
	\label{standardGR}
	e^{-\mu(r)}
	\equiv
	1+\frac{\kappa\,}{r}\int_0^r x^2\,T_0^{\, 0}(x)\, dx
	=
	1-\frac{2\,m(r)}{r}
\end{equation}
is the standard general relativity expression containing the Misner-Sharp mass function $m=m(r)$.
Adding the source $\theta_{\mu\nu}$ results in the GD of the metric~\eqref{pfmetric},
namely
\begin{eqnarray}
	\label{gd1}
	\xi 
	&\rightarrow &
	\nu\,=\,\xi+\alpha\,g
	\\
	\label{gd2}
	e^{-\mu} 
	&\rightarrow &
	e^{-\lambda}=e^{-\mu}+\alpha\,f
	\ , 
\end{eqnarray}
where $f$ and $g$ are respectively the geometric deformations for the radial and temporal metric
components parameterised by $\alpha$.~\footnote{We emphasize that Eqs.~\eqref{gd1} and~\eqref{gd2}
do not represent a coordinate transformation.}
By means of Eqs.~(\ref{gd1}) and (\ref{gd2}), the Einstein equations~\eqref{ec1}-\eqref{ec3}
are separated in two sets:
\begin{itemize}
	\item
	One is given by the standard Einstein field equations for the metric~\eqref{pfmetric} sourced by the
	energy-momentum tensor $T_{\mu\nu}$, that is
\begin{eqnarray}
	\label{ec1pf}
	&&
	\kappa\,\epsilon
	=\frac 1{r^2}
	-
	e^{-\mu }\left( \frac1{r^2}-\frac{\mu'}r\right)\ ,
	\\
	&&
	\label{ec2pf}
	\kappa\,
	\,p_r
	=
	-\frac 1{r^2}
	+
	e^{-\mu}\left( \frac 1{r^2}+\frac{\xi'}r\right)\ ,
	\\
	&&
	\label{ec3pf}
	\kappa\,
	\strut\displaystyle
	\,p_\theta
	=
	\frac {e^{-\mu }}{4}
	\left(2\xi''+\xi'^2-\mu'\xi'
	+2\,\frac{\xi'-\mu'}r\right)
	\ .
\end{eqnarray}
\item The second set contains the source $\theta_{\mu\nu}$ and reads
\begin{eqnarray}
	\label{ec1d}
	\kappa\,{\cal E}
	&=&
	-\frac{\alpha\,f}{r^2}
	-\frac{\alpha\,f'}{r}\ ,
	\\
	\label{ec2d}
	\kappa\,{\cal P}_r
	-\alpha\,Z_1
	&=&
	\alpha\,f\left(\frac{1}{r^2}+\frac{\nu'}{r}\right)
	\\
	\label{ec3d}
	\kappa\,{\cal P}_\theta
	-\alpha\,Z_2
	&=&
	\frac{\alpha\,f}{4}\left(2\,\nu''+\nu'^2+2\frac{\nu'}{r}\right)
	\nonumber
	\\
	&&
	\frac{\alpha\,f'}{4}\left(\nu'+\frac{2}{r}\right)
	\ ,
\end{eqnarray}
where 
\begin{eqnarray}
	\label{Z1}
	Z_1
	&=&
	\frac{e^{-\mu}\,g'}{r}
	\\
	\label{Z2}
	4\,Z_2&=&e^{-\mu}\left(2g''+\alpha\,g'^2+\frac{2\,g'}{r}+2\,g'\xi'-\mu'g'\right)
	\ .
\end{eqnarray}
Of course, the tensor $\theta_{\mu\nu}$ vanishes when the deformations vanish ($f=g=0$). 
\end{itemize}
Finally, the conservation equation~\eqref{dT0} yields
\begin{eqnarray}
	\label{con22}
	\nabla_\sigma\,T^{\sigma}_{\ \nu}=
	-
	\frac{\alpha\,g'}{2}\left(\epsilon+p_r\right)\delta^{\sigma}_{\ \nu}
	=
	-\nabla_\sigma\theta^{\sigma}_{\ \nu}
	\ ,
\end{eqnarray}
which explicitly shows the exchange of energy between the gravitational systems
described by~\eqref{ec1pf}-\eqref{ec3pf} and~\eqref{ec1d}-\eqref{ec3d}, respectively.
The interaction will be pure gravitational (no exchange of energy) when
i) there is no temporal deformation ($g=0$) and
ii) for Kerr-Schild spacetimes with $\epsilon=-p_r$.
This result is particularly remarkable since it is exact, without demanding any perturbative
expansion in $f$ or $g$~\cite{Ovalle:2020fuo}.
\section{Hairy black holes}
\label{sec3}
Our strategy to find hairy deformations of spherically symmetric black holes in general relativity is now
straightforward:
we consider the Schwarzschild metric 
\begin{equation}
	e^\xi
	=
	e^{-\mu}
	=
	1-\frac{2\,M}{r}
	\ ,
	\label{schw}
\end{equation}  
which solves Eqs.~\eqref{ec1pf}-\eqref{ec3pf} for $T_{\mu\nu}=0$ as our seed geometry.
We then search for a matter Lagrangian ${\cal L}_{\Theta}$ corresponding to an energy-momentum
tensor $\theta_{\mu\nu}$ which induces GD $f$ and $g$ in Eqs.~\eqref{ec1d}-\eqref{ec3d} such that
the singularity of the seed metric at $r=0$ is removed.
Note that we have a system of three equations and five unknowns, namely $f$, $g$,
${\cal E}$, ${\cal P}_r$ and ${\cal P}_\theta$.
We are therefore free to impose additional conditions.
\subsection{Horizon structure}
First of all, in order to have black holes with a well-defined horizon structure, we need
$e^{\nu(\rh)}=e^{-\lambda(\rh)}=0$, so that $r=\rh$
will be both a killing horizon ($e^{\nu}=0$) and a causal horizon $(e^{-\lambda}=0)$.
A sufficient condition for this feature is that 
\begin{equation}
\label{constr1}
e^{\nu}=e^{-\lambda}
\ .
\end{equation}
A direct consequence of the Einstein equations~\eqref{ec1} and~\eqref{ec2} with Eq~\eqref{constr1}
is that the source must satisfy the equation of state $\tilde{p}_{r}=-\tilde{\epsilon}$.
For $T_{\mu\nu}=0$, this yields
\begin{equation}
{\cal P}_{r}
	=
	-{\cal E}
	\ ,
	\label{schwcontv}
\end{equation}
and only a negative radial pressure is allowed (for positive energy density).
The critical importance of the condition~\eqref{constr1} is further emphasised by noticing that 
the conservation equation~\eqref{con22} with the equation of state~\eqref{schwcontv}
leads to
\begin{eqnarray}
	\label{xxx}
	{\cal P}_r'=
	\frac{2}{r}\left({\cal P}_\theta-{\cal P}_r\right)
	\ .
\end{eqnarray}
This is precisely the equation of hydrostatic equilibrium which prevents the source $\theta_{\mu\nu}$
to collapse into the central singularity of the seed Schwarzschild metric.
\par
Next, by using the condition~\eqref{constr1} and the seed Schwarzschild solution~\eqref{schw}
in Eqs.~\eqref{gd1}-\eqref{gd2}, we obtain
\begin{equation}
\label{fg}
\alpha\,f
=
\left(1-\frac{2\,M}{r}\right)
\left[e^{\alpha\,g(r)}-1\right]
\ .
\end{equation}
Hence the line element~\eqref{metric} becomes
\begin{eqnarray}
\label{hairyBH}
ds^{2}
&=&
-\left(1-\frac{2\,M}{r}\right)h(r)\,dt^{2}
+\left(1-\frac{2\,M}{r}\right)^{-1}\frac{dr^2}{h(r)}
\nonumber
\\
&&
+r^{2}\left(d\theta ^{2}+\sin {}^{2}\theta\, d\phi ^{2}\right)
\ .
\end{eqnarray}
with
\begin{equation}
\label{h}
h=e^{\alpha\,g(r)}
\ ,
\end{equation}
where $g$ is yet to be determined.

We conclude this section by noting that it is also possible to ensure the existence of a well-defined horizon
with a less restrictive condition than that in Eq.~\eqref{constr1}, namely
\begin{equation}
	\label{constr2}
	e^{\nu(r)}
	=
	e^{\Phi(r)}\,e^{-\lambda(r)}
	\ ,
\end{equation}
where $\Phi$ is regular everywhere.
However, with this more general condition, the equation~\eqref{xxx} for hydrostatic equilibrium
becomes
\begin{eqnarray}
	\label{xxxy}
	{\cal P}_r'=-\frac{1}{2}\left(\Phi'-\lambda'\right)\left({\cal E}+{\cal P}_r\right)+
	\frac{2}{r}\left({\cal P}_\theta-{\cal P}_r\right)
	\ ,
\end{eqnarray}
which makes the analysis much more difficult.
Indeed, the condition~\eqref{constr1} [corresponding to $\Phi=0$] ensures that the hairy solution
found below is still a spacetime of the Kerr-Schild class ($g_{tt}\,g_{rr}=-1$)~\cite{kerrchild},
like most of the known black holes.
In this subclass of spacetimes, the field equations are linear,
greatly simplifying any further analysis.
\subsection{Weak energy conditions}
Even though we can expect that classical energy conditions are generically violated in extreme
high-curvature environments, these conditions remain a good guide to build physically relevant
solutions~\cite{Martin-Moruno:2017exc}.
In this work we require that the tensor vacuum $\theta_{\mu\nu}$ satisfies the weak energy condition
\begin{eqnarray}
\label{weak0a}
&&
{\cal E}
\geq
0
\\
&&
{\cal E}+{\cal P}_r
\geq
0
\label{weak0b}
\\
&&
{\cal E}+{\cal P}_\theta
\geq
0
\ .
\label{weak0c}
\end{eqnarray}
Eq.~\eqref{weak0b} holds as a consequence of~\eqref{schwcontv}, while the
conditions~\eqref{weak0a} and~\eqref{weak0c} are respectively written as
\begin{eqnarray}
	\label{G1}
	\kappa\,r^2\,{\cal E}
	&=&
	-\left(r-2\,M\right)h'-h+1
	\geq
	0
	\\
	\label{G2}
	2\left({\cal E}+{\cal P}_\theta\right)
	&=&
	-r\,{\cal E}'
	\geq
	0
	\ .
\end{eqnarray} 
where $h=h(r)$ is defined in Eq.~\eqref{h}.
Eq.~\eqref{G1} is a first-order linear differential inequality for $h$, whose saturation ${\cal E}=0$
consistently yields the seed Schwarzschild solution~\eqref{schw}.
On the other hand, any possible solution which is regular everywhere will satisfy the inequality~\eqref{G1}
with strictly positive energy density ${\cal E}$ which further decreases monotonously from the origin $r=0$ outwards
in order to also satisfy Eq.~\eqref{G2}, namely ${\cal E}'<0$.
A simple case satisfying all of these conditions is given by
\begin{equation}
\label{G}
\kappa\,{\cal E}
=
\frac{\alpha}{\ell^2}\,e^{-r/\ell}
\ ,
\end{equation}
where $\ell$ is a constant with dimensions of a length.
Notice that $\alpha$ is introduced in Eq.~\eqref{G} in order to recover the seed vacuum solution~\eqref{schw}
for $\alpha\to 0$.
\subsection{Regular spacetime metric}
Using the expression~\eqref{G} in Eq.~\eqref{G1}, we find
\begin{eqnarray}
	\label{strongg}
	h
	=
	\frac{c_1+r}{r-2\,M}
	+
	\frac{\alpha\,e^{-r/\ell}}{r-2\,M}
	\left(2\,\ell+2\,r+\frac{r^2}{\ell}\right)
	\ ,
\end{eqnarray}
where the constant $c_1$ is also a length.
From Eq.~\eqref{strongg} we then obtain the asymptotically flat metric functions
\begin{equation}
	\label{weakBH}
	e^{\nu}
	=
	e^{-\lambda}
	=
	1+\frac{c_1}{r}+\alpha\,e^{-r/\ell}\left(2+\frac{2\,\ell}{r}+\frac{r}{\ell}\right)
	\ .
\end{equation}
Notice that the seed mass $M$ does not appear in the solution~\eqref{weakBH} and the ADM mass
is instead given by ${\cal M}=-c_1/2$. 
Moreover, for $r\sim 0$, one has
\begin{equation}
	\label{zero}
	e^{\nu}
	=
	e^{-\lambda}
	\simeq
	1+\frac{c_1+2\,\alpha\,\ell}{r}-\frac{\alpha\,r^2}{3\,\ell^2}
	\ .
\end{equation}
Hence, the absence of central singularity requires $c_1=-2\,\alpha\,\ell$, which in turn leads to
\begin{equation}
	\label{reg}
	\alpha\,\ell
	=
	{\cal M}
	\ ,
\end{equation}
so that the region around the center is a de~Sitter spacetime with an effective cosmological constant
$\Lambda=\alpha/\ell^{2}$.
Finally, we write the metric~\eqref{weakBH} in terms of the ADM mass~\eqref{reg} as 
\begin{equation}
	\label{weakBH2}
	e^{\nu}
	=
	e^{-\lambda}
	=
	1-\frac{2\,{\cal M}}{r}+\frac{e^{-\alpha\,r/{\cal M}}}{r\,{\cal M}}
	\left(\alpha^2r^2+2\,{\cal M}\alpha\,r+2\,{\cal M}^2\right)
	\ .
\end{equation}
We can see from Eq.~\eqref{weakBH} that the Schwarzschild solution
is recovered for $\alpha\to 0$ (with $c_1=-2\,{\cal M}$).
However, this changes radically after imposing the regularity condition in Eq.~\eqref{reg}.
Notice that we now obtain the Minkowski spacetime for $\alpha\to 0$,
while we recover the Schwarzschild solution for $\alpha\to\infty$.
In fact, the mass function now reads
\begin{equation}
\label{massx}
\tilde m
=
{\cal M}
-\frac{e^{-\alpha\,r/{\cal M}}}{2{\cal M}}\left(\alpha^2r^2+2\,{\cal M}\alpha\,r+2\,{\cal M}^2\right)
\end{equation}
and we further notice that, for $r\to 0$, it vanishes as
\begin{equation}
\tilde m
\simeq
\frac{\alpha^3\,r^3}{6\,{\cal M}^2}
\ .
\end{equation}
This shows that the GD deformation of the seed Schwarzschild metric~\eqref{schw}
is a formal procedure that effectively helps to find new BH solutions with prescribed
physical properties, but which cannot necessarily be obtained by physical deformations
of the seed metric.
\par
\begin{figure*}[ht!]
	\centering
	\includegraphics[width=8cm]{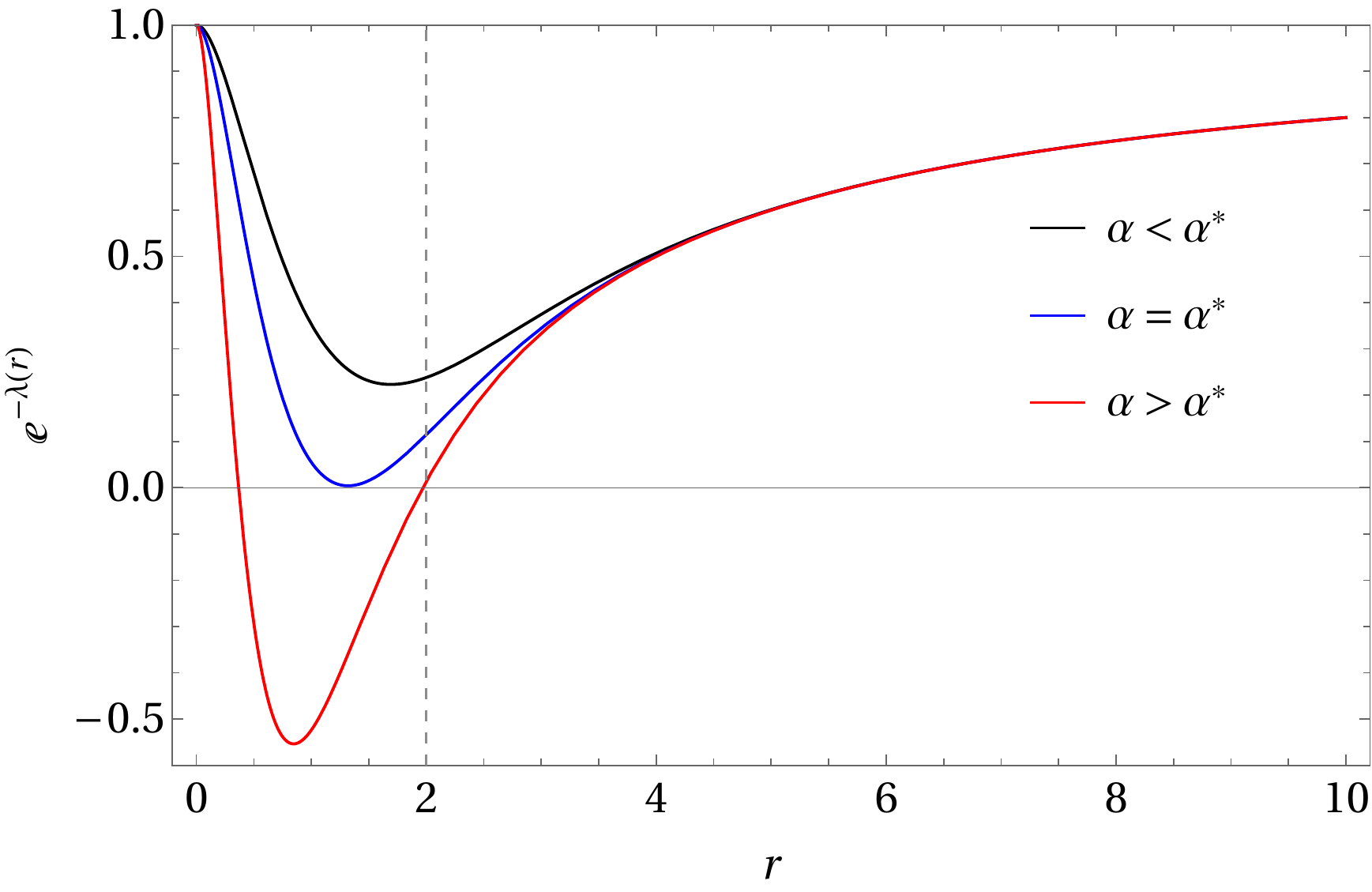}
	$\quad$
	\includegraphics[width=8cm]{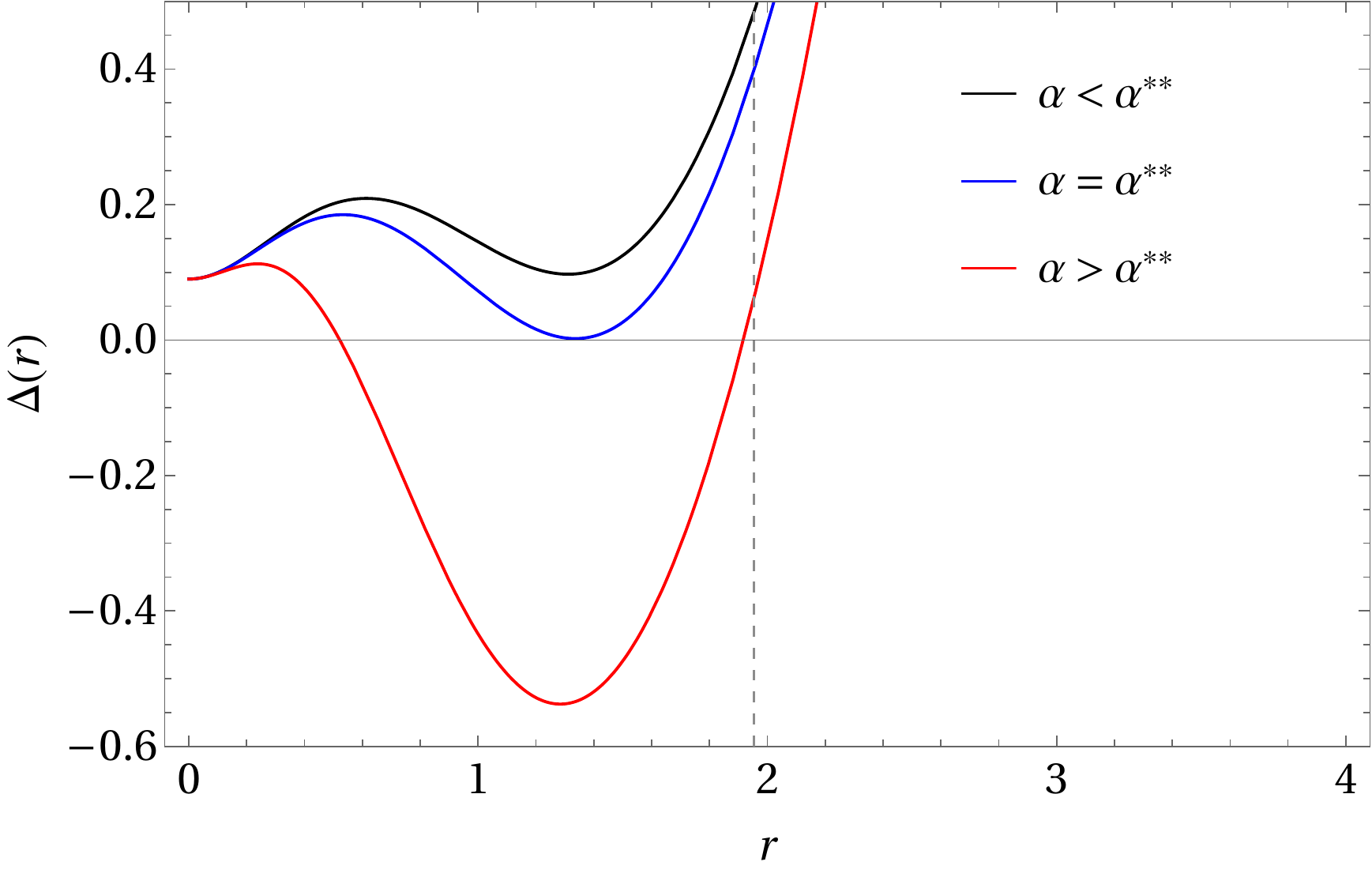}
	\caption{Metric functions for spherically symmetric solutions (left panel) and axially symmetric solutions (right panel).
	Each solution shows three cases, {\em i.e.}, no horizon, extremal black hole and black hole with two horizons.
	The extremal cases are given respectively by $\alpha^*\simeq 2.56$ and $\alpha^{**}\simeq 2.71$.
	When $\alpha\gg\alpha^{*}$ we reproduce the Schwarzschild solution, in agreement with Eq.~\eqref{weakBH2}.
	Likewise, when $\alpha\gg\alpha^{**}$ we obtain the Kerr horizon, in agreement with Eqs.~\eqref{kerrex}
	and~\eqref{massx}.
	All quantities are shown for ${\cal M}=1$.}
	\label{fig1}
\end{figure*}

Possible horizons are found from solutions $\rh=\rh({\cal M},\,\alpha)$ of
\begin{equation}
	\label{horizon}
e^{{-\lambda}(\rh)}=0
\ .
\end{equation}
A standard analysis of Eq.~\eqref{horizon} shows an extremal case for $\alpha=\alpha^*$,
with no zeros for $\alpha<\alpha^*$ and two zeros for $\alpha>\alpha^*$.
These two solutions are the Cauchy and event horizons, as displayed in Fig.~\ref{fig1}
(left panel), where we see the metric for three different cases, {\em i.e.}, without an
event horizon, the extremal configuration, and a black hole.
\par
The scalar curvature is given by
\begin{equation}
	\label{curv}
	R
	=
	\alpha^3
	\left(\frac{4\,{\cal M}-\alpha\,r}{{\cal M}^3}\right)e^{-\alpha\,r/{\cal M}}
\end{equation}
and Ricci squared reads
\begin{equation}
	R_{\mu\nu}\,R^{\mu\nu}
	=
	\alpha^6
	\left(\frac{8\,{\cal M}^2-4\,\alpha\,{r}{\cal M}+\alpha^2\,r^2}{2{\cal M}^6}\right)e^{-2\alpha{r}/{\cal M}}\ ,
	\label{curvx}
\end{equation}
while the complete expression of the Kretschmann scalar $R_{\mu\nu\rho\sigma}\,R^{\mu\nu\rho\sigma}$ is too involved
for displaying.
For $r\to 0$, it behaves as
\begin{equation}
\label{eq:K}
	R_{\mu\nu\rho\sigma}\,R^{\mu\nu\rho\sigma}
	\simeq
	\frac{8\,\alpha^6}{3\,{\cal M}^4}-\frac{20\,\alpha^7\,r}{3\,{\cal M}^5}+\frac{35\,\alpha^8\,r^2}{4\,{\cal M}^6}
	\ ,
\end{equation}
so we conclude that the solution has no curvature singularities.
\par
\begin{figure}[t]
	\center
	\includegraphics[width=8cm]{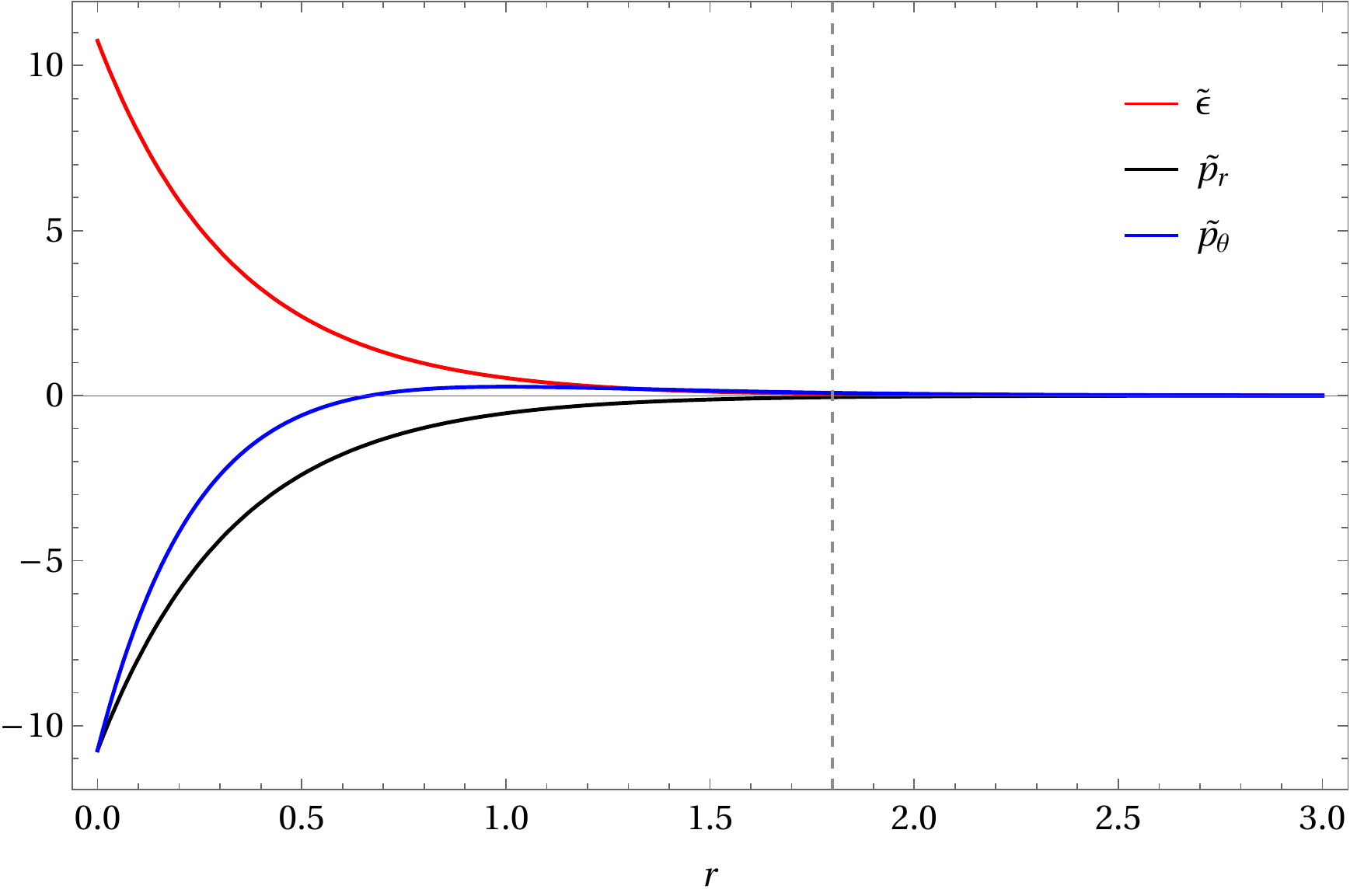}
	\\
	\caption{Source terms $\{\tilde{\epsilon},\,\tilde{p}_r,\,\tilde{p}_\theta\}\times 10$ in the spherically symmetric case
	for $\alpha=3$. 
	The vertical line shows the event horizon $\rh\sim\,1.8$. 
	All quantities shown for ${\cal M}=1$.}
	\label{fig2}      
\end{figure}

Finally, the source generating the metric functions~\eqref{weakBH2} has the effective density~\eqref{G} and
an effective tangential pressure
\begin{equation}
	{\cal P}_{\theta}
	=
	\alpha^3
	\left(\frac{\alpha\,r-2\,{\cal M}}{2\,\kappa\,{\cal M}^3}\right)e^{-\alpha\,r/{\cal M}}\ .
	\label{prestrong}
\end{equation}
We further have
\begin{equation}
\label{weak5}
{\cal E}+{\cal P}_{\theta}
=
\frac{\alpha^4\,r}{2\,\kappa\,{\cal M}^3}e^{-\alpha\,r/{\cal M}}
\ ,
\end{equation}
and the WEC indeed holds for $r\geq 0$, as displayed in Fig.~\ref{fig2}, where we see that 
the vacuum is approached very quickly outside the event horizon $r=\rh$. 
Also notice that, in agreement with Eq.~\eqref{xxx}, the fluid experiences a pull towards the center
from the radial pressure gradient ${\cal P}_r'>0$, which is precisely canceled by a gravitational repulsion
caused by the pressure anisotropy $\Pi={\cal P}_\theta-{\cal P}_r$.
%

\section{Axially symmetric case}
\label{sec4}
In order to build the rotating version of the metric in Eq.~\eqref{weakBH2}, we follow the strategy described
in Ref.~\cite{Contreras:2021yxe}
(see also Refs.~\cite{Burinskii:2001bq,Dymnikova:2006wn,Smailagic:2010nv,Bambi:2013ufa,Azreg-Ainou:2014nra,Dymnikova:2016nlb}).
This simply amounts to consider the general Kerr-Schild metric in Boyer-Lindquist coordinates, namely, the Gurses-Gursey metric~\cite{Gurses:1975vu}
\begin{eqnarray}
	\label{kerrex}
	ds^{2}
	&=&
	-\left[1-\frac{2\,r\,{\tilde{m}}(r)}{{\rho}^2}\right]
	dt^{2}
	-
	\frac{4\, {a}\, r\,{\tilde{m}}(r)\, \sin^{2}\theta}{{\rho}^{2}}
	\,dt\,d\phi
	\nonumber
	\\
	&&
	+
	\frac{{\rho}^{2}}{{\Delta}}\,dr^{2}
	+
	{\rho}^{2}\,d\theta^{2}
	+
	\frac{{\Sigma}\, \sin^{2}\theta}{{\rho}^{2}}\,d\phi^{2}
	\ ,
\end{eqnarray}
with
\begin{eqnarray}
	{\rho}^2
	&=&
	r^2+{a}^{2}\cos^{2}\theta
	\label{f0}
	\\
	{\Delta}
	& = &
	r^2-2\,r\,{\tilde{m}}(r)
	+{a}^{2}
	\label{f2}
	\\
	{\Sigma}
	& = &
	\left(r^{2}+{a}^{2}\right)^{2}
	-{\Delta}\, a^2\sin^{2}\theta
	\label{f3}\\
	{a} & = & {J}/{\cal M}
	\ ,
\end{eqnarray}
where $\tilde m$ is the mass function of our reference spherically symmetric metric~\eqref{weakBH2}
given in Eq.~\eqref{massx},
${J}$ is the angular momentum and ${\cal M}=\tilde m(r\to\infty)$ the total mass of the system.
Clearly, Eq.~\eqref{kerrex} reduces to the Kerr solution when $\tilde m=M$ and, as remarked
in Ref.~\cite{Contreras:2021yxe}, it is not necessary to resort to the Newman-Janis algorithm.
\par
The line-element~\eqref{kerrex} contains two potential singularities, namely, when $\rho=0$ or $\Delta=0$.
The case $\rho=0$ is the ring singularity of the Kerr solution, and it is a physical singularity
which occurs at $\theta=\pi/2$ and $r=0$.
The curvature scalar of the line element~\eqref{kerrex} reads
\begin{equation}
	\label{R}
	R
	=
	\frac{2\,(2\,\tilde{m}'+r\,\tilde{m}'')}{\rho^2}\ ,
\end{equation}
which, for the mass function~\eqref{massx}, yields
\begin{equation}
	\label{R2}
	R
	=
	\frac{\alpha^3\,r^2\,e^{-\alpha\,r/{\cal M}}}{\rho^2\,{\cal M}^3}\left(4{\cal M}-\alpha\,r\right)
	\ .
\end{equation}
We see that the expression in Eq.~\eqref{R2} is regular for $r=0$ and $\theta=\pi/2$.
The Ricci squared $R_{\mu\nu}\,R^{\mu\nu}\vert_{\theta=\pi/2}$ has the same regular form
displayed in Eq.~\eqref{curvx}, while the Kretschmann scalar for $r\sim 0$ and $\theta = \pi/2$ reads as in \eqref{eq:K}.
We can conclude that our rotating solution is free of physical singularities. 
\par
As usual, the region $\Delta=0$ represents a coordinate singularity that indicates the existence of horizons,
defined by
\begin{equation}
	\label{Delta}
	\Delta(\rh)=\rh^2-2\,\rh\,{\tilde{m}}(\rh)
	+{a}^{2}=0
	\ .
\end{equation}
The expression in Eq.~\eqref{Delta} reveals an extremal case for $\alpha=\alpha^{**}$, with no zeros
for $\alpha<\alpha^{**}$ and always two zeros for $\alpha>\alpha^{**}$.
These two solutions are the Cauchy and event horizons, as displayed in Fig.~\ref{fig1} (right panel),
where we again observe the same three cases.
\par
Finally, the energy-momentum tensor $\theta_{\mu\nu}$ generating the metric~(\ref{kerrex})
is given by
\begin{eqnarray}\label{tmunu}
	\theta^{\mu\nu}
	=
	\tilde{\epsilon}\, {u}^{\mu}\,{u}^{\nu}
	+\tilde{p}_{r}\,{l}^{\mu}\,{l}^{\nu}
	+\tilde{p}_{\theta}\,{n}^{\mu}\,{n}^{\nu}
	+\tilde{p}_{\phi}\,{m}^{\mu}\,{m}^{\nu}
	\ ,
\end{eqnarray}
where the orthonormal tetrad reads~\cite{Gurses:1975vu}~\footnote{Note that $+\Delta$ refers to the regions
outside the event horizon and inside the Cauchy horizon, while $-\Delta$ refers to the region between the two horizons.}
\begin{eqnarray}
	\label{tetrada}
	{u}^{\mu}
	&=&
	\frac{(r^{2}+{a}^{2})\delta^\mu_0+a\,\delta^\mu_3}{\sqrt{\pm\Delta\rho^{2}}}
	\ ,
	\qquad
	{l}^{\mu}
	=
	\sqrt{\frac{\pm\Delta}{\rho^{2}}}\,\delta^\mu_1
	\label{2}
	\nonumber
	\\
	{n}^{\mu}
	&=&
	\frac{1}{\sqrt{\rho^{2}}}\,\delta^\mu_2
	\ ,
	\qquad
	{m}^{\mu}
	=
	-\frac{{a}\sin^{2}\theta\,\delta^\mu_0+\delta^\mu_3}{\sqrt{\rho^{2}}\sin\theta}
	\ ,
	\label{4}
\end{eqnarray}
and the energy density $\tilde{\epsilon}$ and pressures $\tilde{p}_r$, $\tilde{p}_\theta$ and $\tilde{p}_\phi$
are given by
\begin{eqnarray}
	\label{energyax}
	{ \kappa \,} \tilde{\epsilon}
	&=&
	- { \kappa \,} \tilde{p}_{r}
	=
	\frac{2\,r^2}{\rho ^4}\, \tilde{m}'\ ,
	\\
	\label{pressuresax}
	{ \kappa \,} \tilde{p}_{\theta}
	&=&
	{ \kappa \,} \tilde{p}_{\phi}
	=
	-\frac{r }{\rho ^2}\,\tilde{m}''
	+\frac{2\left(r^2-\rho^2\right)}{\rho ^4} \,\tilde{m}'
	\ .
\end{eqnarray}
\section{Conclusions}
\label{con}
The appearance of singularities as the final result of the gravitational collapse is a well-known
prediction in the framework of GR.
One way to avoid such singularities is to introduce non-collapsing matter.
This has been our strategy, introducing what we generically call tensor vacuum and is
explicitly described by the expression~\eqref{xxx}.
This allows us to construct both static and stationary regular hairy BHs,
whose hair is parametrised by a parameter $\alpha$ with a clear physical interpretation,
as we can read from solutions~\eqref{weakBH2} and~\eqref{kerrex}, that is
\begin{itemize}
	\item $\alpha\to\,0\Rightarrow$ Minkowski
	\item $\alpha\to\,\infty\Rightarrow$ Schwarzschild (static case)
	\item $\alpha\to\,\infty\Rightarrow$ Kerr (stationary case)
\end{itemize}
We can therefore interpret the BH hair as the source which deformas the Minkowski vacuum
(total absence of matter and gravity) with a maximum deformation corresponding precisely
to the Schwarzschild solution for the static case, and the Kerr solution for the stationary case.
The formation of BHs occurs beyond the critical values $\alpha^*$ and $\alpha^{**}$, respectively,
as we show in Fig.~\ref{fig1}.
\par
In the present work, we do not make any attempt at postulating the action from which
to derive the solution~\eqref{weakBH2}.
However, we remark that, by means of the $P$-dual formalism~\cite{Salazar:1987ap,Ayon-Beato:1998hmi},
it is always possible to find a Lagrangian ${\cal L}$ associated with a theory
which results in a given energy-momentum $\theta_{\mu\nu}$ producing the geometric
deformation $f$ and $g$, with mass function 
\begin{equation}
\label{standardGR2}
\tilde{m}(r)=
\frac{{ \kappa}}{2}\int_0^r x^2\,\theta_0^{\, 0}(x)\, dx
\ .
\end{equation}
In this approach, the total action reads
\begin{equation}
\label{ngt}
S_{\rm G}=\int\left[\frac{R}{2\kappa}+{\cal L}(F)\right]\sqrt{-g}\,d^4\,x
\ ,
\end{equation}
where ${\cal L}(F)$
can be obtained by means of the $P$-dual formalism and reads (see Appendix~\ref{A:L}) 
\begin{eqnarray}
	\label{Lp}
	{\cal L}(P)
	=\frac{\alpha^3}{2\,\kappa\,{\cal M}^3}\left[{\xi(P)}-2\,{\cal M}\right]	\,e^{-\xi(P)/{\cal M}}
	\ ,
\end{eqnarray}
with
\begin{eqnarray}
	\xi(P)
	=\alpha\left(\frac{-2\,\alpha^2}{\kappa^2\,P}\right)^{1/4}
	\ .
\end{eqnarray}
\par
We conclude by mentioning that some aspects of the presented solutions should be analyzed more in depth,
like their stability, observational consequences, time-dependent formation and evaporation.
As for the feasibility of removing the Cauchy horizon without risking the regularity of the
solution, it is likely that one should resort to quantum physics~\cite{Casadio:2022ndh,Casadio:2023iqt}.
\subsection*{Acknowledgments}
J.O.~is partially supported by ANID FONDECYT grant No.~1210041. R.C.~is partially supported by the INFN grant FLAG.
The work of R.C.~and A.G.~has also been carried out in the framework of activities of the
National Group of Mathematical Physics (GNFM, INdAM).
\appendix
\section{Additional Lagrangian}
\label{A:L}
\setcounter{equation}{0}
In order to specify the theory encoded by ${\cal L}(F)$ in Eq.~\eqref{ngt}, we identify
\begin{eqnarray}
	\label{tmunu}
	\theta_{\mu\nu}
	=
	-
	{\cal L}_{F}\,F_{\mu\alpha}\,F^{\alpha}{} _{\nu}
	-{\cal L}(F)\,g_{\mu\nu}
	\ ,
\end{eqnarray}
where 
\begin{equation}
	F
	=
	\frac{1}{4}\,F_{\mu\nu}\,F^{\mu\nu}
	\quad
	{\rm and}
	\quad 
	{\cal L}_F=\frac{d{\cal L}}{dF}
\end{equation}
is a non-linear Maxwell representation of the theory.
At this stage we emphasize that this theory is not necessarily a nonlinear electrodynamics,
in the sense that the charge, or primary hair generating it, is not necessarily an electric charge.
\par
In the static spherically symmetric case, we have
\begin{equation}
	\label{fs}
	F_{\mu\nu}
	=
	E(r)\,\left(\delta^0_\mu\,\delta^1_\nu-\delta^1_\mu\,\delta^0_\nu\right)
	\ ,
\end{equation}
where $E$ is the ``electric field''.
Using Eqs.~\eqref{constr1} and~\eqref{tmunu}-\eqref{fs} in the Einstein
equations~\eqref{ec1} and~\eqref{ec2}, we obtain
\begin{eqnarray}
	-\frac{2}{r^{2}}\frac{d\tilde{m}}{dr}
	&=&
	\kappa\,
	\left[{\cal L}(F)+E^{2}\,{\cal L}_{F}\right]
	\label{1}
	\\
	-\frac{1}{r}\frac{d^{2}\tilde{m}}{dr^{2}}
	&=&
	\kappa\, {\cal L}(F)
	\ ,
	\label{2}
\end{eqnarray}
where $\tilde{m}$ is the Misner-Sharp mass function given in Eq.~\eqref{massx}.
The corresponding conservation equation~\eqref{dT0} reads $\nabla_{\mu}({\cal L}_{F}\,F^{\mu\nu})=0$
and leads to
\begin{eqnarray}
	\label{electrico}
	E\,{\cal L}_{F}
	=
	-\frac{2\,\alpha}{\kappa\,r^{2}}
	\ .
	\label{3}
\end{eqnarray}
Notice that in Eq.~\eqref{3} we use the parameter $\alpha$ as the charge generating the field.
On subtracting~\eqref{1} from Eq.~\eqref{2}, we obtain
\begin{eqnarray}
	\label{aaa}
	r\,\frac{d}{dr}\left(\frac{1}{r^{2}}\frac{d\tilde{m}}{dr}\right)
	=
	\kappa\, E^{2}\,{\cal L}_{F}
	\ .
\end{eqnarray}
Finally, combining Eqs.~\eqref{electrico} and~\eqref{aaa} we obtain
\begin{eqnarray}
	E
	=
	-\frac{r^{3}}{2\,\alpha}\,\frac{d}{dr}\left(\frac{1}{r^{2}}\frac{d\tilde{m}}{dr}\right)
	\ .
	\label{el}
\end{eqnarray}
Hence, the explicit form of the field~\eqref{el} generating the black hole solution
described by the metric~\eqref{weakBH2} reads
\begin{eqnarray}
	E
	=
	\frac{\alpha^3}{4\,{\cal M}^3}\,r^3\,e^{-\alpha\,r/{\cal M}}
	\ .
\end{eqnarray}
Notice that by using Eqs.~\eqref{1}-\eqref{3} we cannot obtain the explicit form ${\cal L}={\cal L}(F)$. In order to find the Lagrangian ${\cal L}$ of the underlying theory,
we will use the $P$-dual formalism~\cite{Salazar:1987ap,Ayon-Beato:1998hmi},
which is based on the Legendre transformation
\begin{eqnarray}
	H
	=
	2\,F\,{\cal L}_{F}-{\cal L}
	\ ,
\end{eqnarray}
where $H$ represents the Hamiltonian in the dual formulation.
Now, defining $P_{\mu\nu}=L_{F}\,{\cal L}_{\mu\nu}$, it is straightforward to see that $H$
is a function of $P=P_{\mu\nu}\,P^{\mu\nu}/4$ so that we can write
(for all the details, see Ref.~\cite{Salazar:1987ap})
\begin{eqnarray}
\label{LP}
	{\cal L}
	=
	2\,P\, H_{P}-H
	\ ,
\end{eqnarray}
where $H_P$ denotes the derivative of $H$ with respect to its argument $P$.
In terms of $H$ the energy--momentum tensor reads
\begin{eqnarray}
\label{eP}
	\theta_{\mu\nu}
	=
	-H_{P}\,P_{\mu\alpha}P^{\alpha}{}_{\nu}
	-g_{\mu\nu}\left(2\,P\,H_{P}-H\right)
	\ .
	\end{eqnarray}
Since we are interested in a static and spherically symmetric case, we take
$P_{\mu\nu}=(\delta^{0}_{\mu}\,\delta^{1}_{\nu}-\delta^{1}_{\mu}\,\delta^{0}_{\nu})D(r)$,
where $D$ is the dual field and $H$ is given by
\begin{eqnarray}
\label{HH}
	H
	=
	\frac{2}{\kappa\, r^{2}}\,\frac{d\tilde{m}}{dr}
	\ .
\end{eqnarray}
Since $P={\cal L}_{F}^{2}\,F=-{\cal L}_{F}^{2}\,E^{2}/2$, we obtain
\begin{eqnarray}
	\label{Pf}
	P
	=
	-\frac{2\,\alpha^{2}}{\kappa^2\,r^{4}}
	\ .
\end{eqnarray}
Finally, by using Eqs.~\eqref{massx},~\eqref{HH} and~\eqref{Pf} in Eq.~\eqref{LP}, 
we obtain the Lagrangian in Eq.~\eqref{Lp}.

%

%
%
\end{document}